\def\year{2017}\relax
\documentclass[letterpaper]{article}
\usepackage{aaai17}
\usepackage{times}
\usepackage{helvet}
\usepackage{courier}
\usepackage{graphicx}
\usepackage{amssymb,amsmath,bm}
\usepackage{subfigure}
\usepackage{tabularx,threeparttable}
\usepackage{booktabs}
\usepackage{array,multirow}
\usepackage{color}

\begin{document}
%
\title{Joint Segmentation and Landmark Localization of Fetal Femur in \\Ultrasound Volumes}
\author{
	Xu Wang$^{1,2\dagger}$,
	Xin Yang$^{3\dagger}$, 
	Haoran Dou$^{1,2}$,
	Shengli Li$^{4}$,
	Pheng-Ann Heng$^{3}$,
	Dong Ni$^{*1,2}$\\
	$^{1}$National-Regional Key Technology Engineering Laboratory for Medical Ultrasound, \\
	School of Biomedical Engineering, Shenzhen University, China\\
	$^{2}$Medical UltraSound Image Computing (MUSIC) Lab, Shenzhen University, China\\
	$^{3}$Department of Computer Science and Engineering, The Chinese University of Hong Kong, Hong Kong, China\\
	$^{4}$Department of Ultrasound, Affiliated Shenzhen Maternal and Child Healthcare, \\Hospital of Nanfang Medical University, Shenzhen, China\\
	\textit{*nidong@szu.edu.cn}
}

\maketitle
\begin{abstract}
	Volumetric ultrasound has great potentials in promoting prenatal examinations. Automated solutions are highly desired to efficiently and effectively analyze the massive volumes. Segmentation and landmark localization are two key techniques in making the quantitative evaluation of prenatal ultrasound volumes available in clinic. However, both tasks are non-trivial when considering the poor image quality, boundary ambiguity and anatomical variations in volumetric ultrasound. In this paper, we propose an effective framework for simultaneous segmentation and landmark localization in prenatal ultrasound volumes. The proposed framework has two branches where informative cues of segmentation and landmark localization can be propagated bidirectionally to benefit both tasks. As landmark localization tends to suffer from false positives, we propose a distance based loss to suppress the noise and thus enhance the localization map and in turn the segmentation. Finally, we further leverage an adversarial module to emphasize the correspondence between segmentation and landmark localization. Extensively validated on a volumetric ultrasound dataset of fetal femur, our proposed framework proves to be a promising solution to facilitate the interpretation of prenatal ultrasound volumes.
\end{abstract}

\section{Introduction}
Characterized with real-time imaging, low-cost and free of radiation, ultrasound imaging is a dominant modality in prenatal examinations. Larger field of view and more freedom for acquisition equip volumetric ultrasound with more compelling potentials than 2D ultrasound for fetal and maternal health evaluation. However, manually conducting quantitative analysis for the massive volumes is time-consuming and challenging in clinical workflow \cite{yang2018towards}. \par

\begin{figure}[htb]
	\begin{minipage}[b]{1.0\linewidth}
		\centering
		\includegraphics[width=1.0\textwidth]{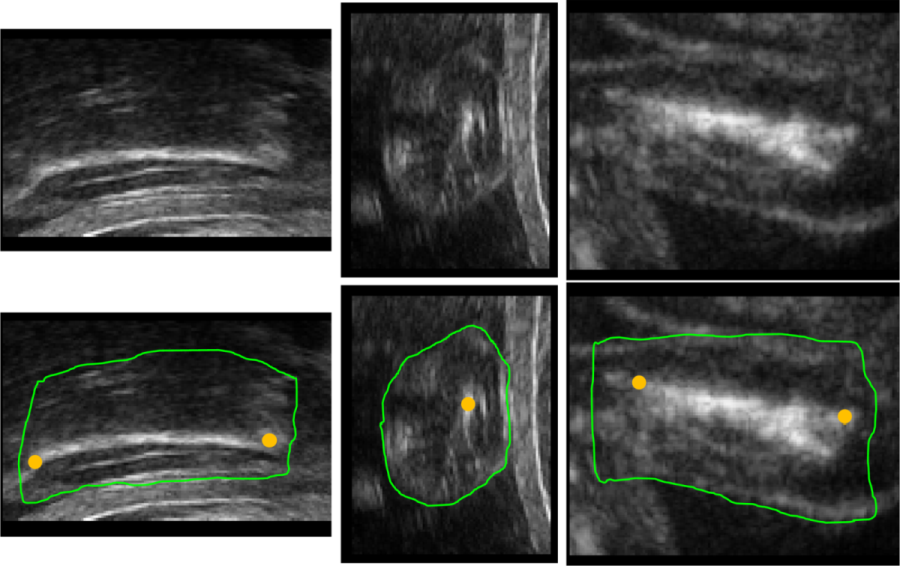}
	\end{minipage}
	\caption{Orthogonal slices of fetal femur in an ultrasound volume. Green curve denotes the segmentation ground truth, yellow dots denote two endpoints.}
	\label{fig:challenges}
\end{figure}

Segmentation and landmark localization are two key techniques for automatic analysis of prenatal ultrasound volumes. Segmentation provides volumetric measurements of fetal and maternal anatomical structures, which may be more comprehensive and accurate than 2D ones for fetal growth evaluation \cite{yang2018towards}. Landmark localization can provide detailed descriptions of anatomical poses and geometries, which are helpful for further applications, such as standard plane detection \cite{li2018standard} and atlas construction \cite{huang2018omni}. However, as the fetal femur shown in Fig. \ref{fig:challenges}, conquering these two tasks is non-trivial. Firstly, volumetric ultrasound is often criticized by its poor image quality, such as speckle noise, acoustic shadow and low resolution. Two tips of the fetal femur are suffering from these factors and thus hard to be localized. Secondly, boundary deficiency and ambiguity often occur as results of shadow occlusion and low contrasts among tissues. Thirdly, the varying pose, shape and size of anatomical structures make it hard for automatic algorithms to capture the appearance variation in ultrasound volumes. \par

Intensive researches for segmentation and landmark localization in ultrasound volumes have been conducted. For segmentation, Feng et al. \cite{Feng_limb}constructed boundary traces to extract fetal limb volume for weight estimation. Namburete et al. \cite{namburete2015learning} built a B-spline surface model to parameterize fetal skull volume. These shape models are robust against artifacts but have limited deformation ranges and are initialization-dependent. In \cite{yang2018towards}, a deep neural network based method was proposed for prenatal ultrasound volume segmentation. The result is promising but it is patch based and suffers from losing global shape information. Cerrolaza et al. \cite{cerrolaza20183d}  proposed to reconstruct fetal skull with conditional generative neural networks. However, the reconstruction may be limited to rigid objects with little deformation. For landmark detection, Huang et al. \cite{huang2018vp} proposed to localize several fetal brain structures by projecting the 3D problem into a 2D task and solved it with deep networks. In \cite{huang2018omni}, Huang et al. further reformulated the localization as a segmentation task. In \cite{namburete2018fully}, a branched deep network was proposed to segment fetal skull and localize fetal eyes to assist the alignment of fetal brain. However, the dependence between segmentation and landmark localization was not fully exploited in the work. \par

In this paper, we propose an end-to-end deep neural network to simultaneously tackle the segmentation and endpoint localization of fetal femur in prenatal ultrasound volumes. Volume and length of fetal femur have unique importance in fetal weight estimation \cite{Feng_limb}. The proposed framework takes two cross-connected network branches as a backbone, where informative cues of segmentation and landmark localization can be propagated bidirectionally to benefit each other. As femur landmark localization encounters more false positives than segmentation, we propose a distance based loss to exclude unreasonable predictions and thus suppress the noise to enhance the localization map. We find that improving femur landmark localization can in turn help improve the segmentation. Finally, to emphasize the correspondence between segmentation and landmark localization, we further leverage an adversarial module to serve as a constraint in a weakly supervised way. Extensive experiments prove our proposed framework as a promising solution to improve the efficiency and accuracy in analyzing fetal femur in ultrasound volumes. \par

\section{Methodology}
\begin{figure}[htb]
	\begin{minipage}[b]{1.0\linewidth}
		\centering
		\includegraphics[width=1.0\textwidth]{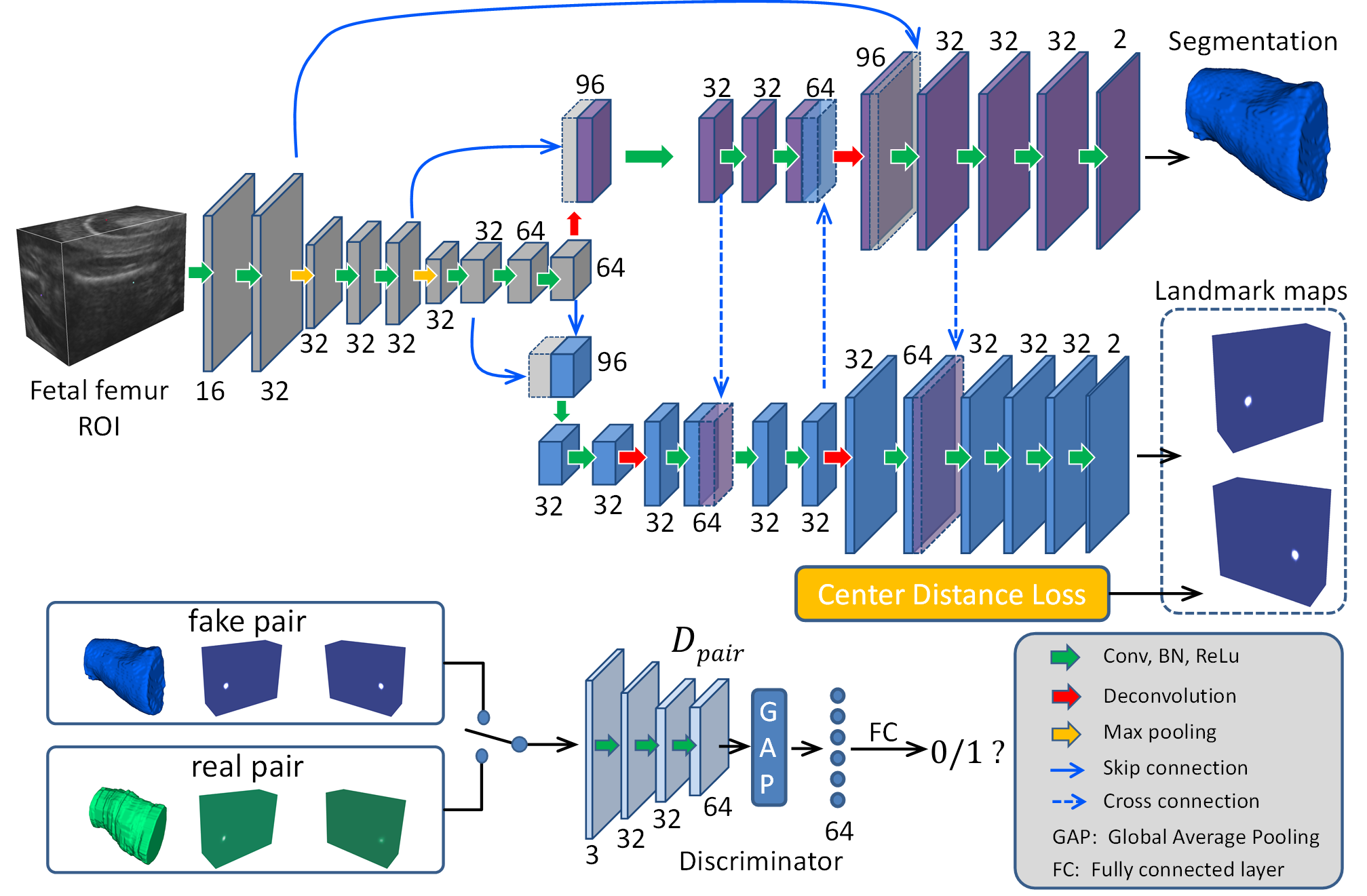}
	\end{minipage}
	\caption{Our framework for simultaneous segmentation (upper branch) and landmark localization (lower branch). Discriminator forces the segmentation and localization to match each other. Digits for number of feature maps.}
	\label{fig:framework}
\end{figure}

Our proposed framework is shown in Fig.~\ref{fig:framework}. Fetal femur ROI is firstly detected in the whole ultrasound volume. Segmentation and landmark localization branches receive the common features of ROI extracted by the shared layers, and then generate task-specific descriptors. Cross connections enable the communication between two branches. Our proposed distance loss is used to improve the landmark localization. Finally, an adversarial discriminator is connected to emphasize the correspondence between two branches. \par

\subsection{ROI Detection Network}
\begin{figure}[htb]
	\begin{minipage}[b]{1.0\linewidth}
		\centering
		\includegraphics[width=1.0\textwidth]{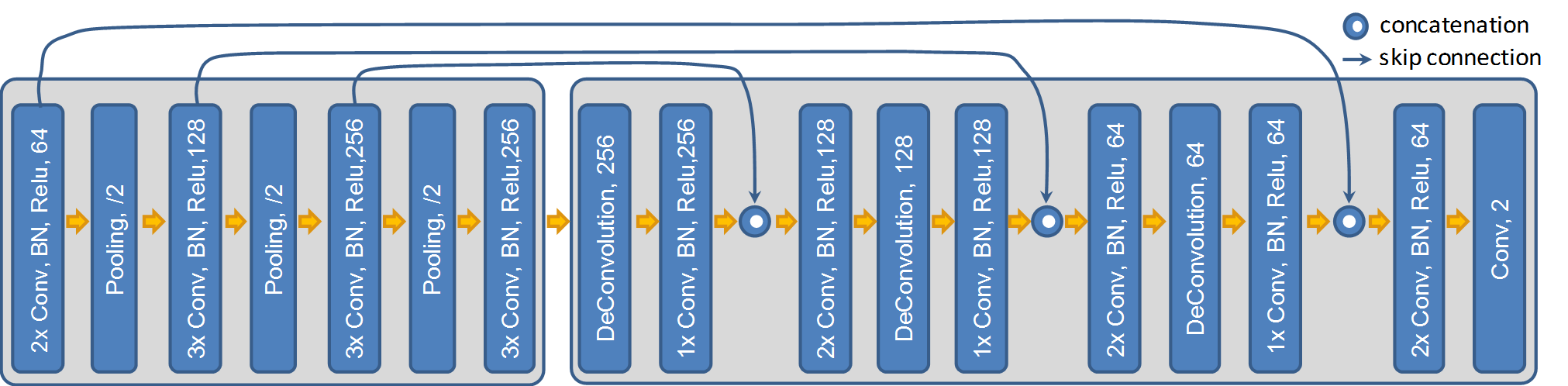}
	\end{minipage}
	\caption{Network details of the ROI detector for fetal femur localization.}
	\label{fig:roi_detector}
\end{figure}

To reduce the searching area, we propose to build a basic U-net \cite{ronneberger2015u} (denoted as \textit{Unet-ROI}) to localize the ROI of fetal femur via segmentation. Limited by GPU memory, \textit{Unet-ROI} is implemented in 2D (detailed layout in Fig. \ref{fig:roi_detector}). 3$\times$3 kernel for convolutional layers (Conv). A batch normalization (BN) layer and rectified linear unit (ReLU) follow the Conv. Segmentation performance of \textit{Unet-ROI} is reported in Table \ref{table:Comparison_seg}. For a testing volume, all the 2D bounding boxes of the segmented femur parts in slices are merged into a 3D bounding box. Our resulted bounding boxes achieve an average $78.1\%$ IoU in localizing the ground truth of femur areas. To completely cover the femur in testing volumes, the 3D bounding box is further augmented with 30 voxels on each dimension to form the final femur ROI. Average size of detected ROIs is 228$\times$124$\times$190. \par

\subsection{Cross-connected Network Architecture}
Receiving the cropped ROI as input, our segmentation and landmark localization branches also follow the U-net design (Fig. \ref{fig:framework}). The encoder path is shared by both branches, which provides basic feature hierarchies of the ROI. For segmentation branch, two long skip connections are established between encoder and decoder paths to enrich the semantic features and improve the segmentation details. Concatenation is used for the skip connections in merging features. Since landmark localization is more difficult than segmentation, we deepens the encoding path of localization branch to get a higher-level understanding of the input ROI. A skip connection is used in localization branch to fuse features before decoding. We use convolution kernels with size of 3$\times$3$\times$3 in all Conv layers. \par

Segmentation branch outputs probability maps of different classes and then the final segmentation, while localization branch outputs independent heatmap for each landmark (Fig. \ref{fig:framework}). Femur landmark location is defined as the point with highest intensity in the heatmap. For the loss function, we take the hybrid loss ($\mathcal{L}_{hybrid}$) for segmentation \cite{yang2017hybrid}. $\mathcal{L}_{hybrid}$ combines weighted cross entropy (wCross) and Dice Similarity Coefficient (DSC) based losses, i.e. $\mathcal{L}_{hybrid}=\mathcal{L}_{wCross}+\theta\mathcal{L}_{DSC}$. $\theta=100$ to balance the scales of two loss components. For the landmark localization, we use the L2-norm regression loss ($\mathcal{L}_{reg}$), which is calculated between the predicted heatmaps ($\mathcal{\hat{H}}$) and Gaussian maps of landmark labels ($\mathcal{H}$). \par

Segmentation and landmark localization should be two closely related tasks, rather than be independently processed. Segmentation of anatomical structures can clearly define the region where landmarks may exist and thus discards a lot of candidates in background, while landmarks can explicitly describe the topology of anatomical structures and adds geometry constraints on segmentation. Sharing intermediate and complementary knowledge between tasks can hence benefit both. In this regard, inspired by \cite{cheng2017segflow}, as shown in Fig. \ref{fig:framework}, we introduce the bidirectional crossed connections between the same semantic levels of two branches. Specifically, two connections convey the feature maps from segmentation branch to localization branch, while one connection from localization branch to segmentation branch. With the training on these two branches to minimize losses, the communicated feature maps can regularize the opposite branch and thus improve both. \par

\subsection{Localization Constraint with Center Distance Loss}
Because of the poor image quality of ultrasound volume, localizing anatomical landmarks in it is non-trivial and suffers from strong false positives. The task for fetal femur becomes more challenging since these two femur endpoints are very hard to be clearly differentiated from each other, especially in the case of lacking context information. For landmark predictions, severe overlap often occurs between the heatmaps of two landmarks. This kind of overlap can degrade the predictions in each channel and thus causes false positives, missed detection or two overlapped landmarks. Therefore, reducing the heatmap overlap and, at the same time, regularizing the distance between landmarks in a reasonable range, are straightforward to improve the localization. Based on these observations, we introduce a Center Distance (CD) based Loss ($\mathcal{L}_{CD}$) as an extra constraint on landmark localization. The formulation of $\mathcal{L}_{CD}$ is defined in Eq. \ref{eq:center_loss},
\begin{gather}
	\label{eq:center_loss}
	\mathcal{L}_{CD} = 1/\parallel M(\mathcal{\hat{H}}_{1})-M(\mathcal{\hat{H}}_{2}) \parallel_{2}^{2},
\end{gather}
where $\mathcal{\hat{H}}_{i}$ is the predicted heatmap of landmark $i$, $M(\cdot)$ is an operator to get the coordinates of the voxel with highest value in heatmap. Minimizing $\mathcal{L}_{CD}$ can enlarge the distance between two heatmap peaks and hence reduces the heatmap overlap. To now, the loss function to train our branched network for both tasks is defined as Eq. \ref{eq:sum_loss}, where $\alpha$, $\beta$ and $\gamma$ are empirically set as 1.0, 0.2 and 0.5 to weight two branches, respectively.
\begin{gather}
	\label{eq:sum_loss}
	\mathcal{L}_{branch} = \alpha\mathcal{L}_{hybrid} + \beta(\mathcal{L}_{reg} + \gamma\mathcal{L}_{CD}).
\end{gather}

\subsection{Refinement with Adversarial Module}
To further emphasize the correspondence between two tasks and force the outputs to match each other, we extend the branched network into an adversarial scheme for refinement. The core of adversarial training scheme is pushing the generator to produce outputs that can fool the discriminator, while discriminator classifies its input as real or fake \cite{kazeminia2018gans}. In our setting (Fig. \ref{fig:framework}), the branched network is the generator, while a 6-layer convolutional network serves as the discriminator ($D_{pair}$). For the discriminator, kernel size is 4 in Conv layers. Stride in first two Conv layers is 2, the rest is 1. Facing with the varying input femur ROI dimension, a global average pooling layer is used before fully connection layer to unify the feature dimension. The generator outputs the concatenation of fetal femur segmentation and landmark heatmaps $(\hat{y}, \mathcal{\hat{H}}_{1}, \mathcal{\hat{H}}_{2})$ and then inputs it to the $D_{pair}$ as a fake pair. The fake pair is then classified by discriminator against the real pair of segmentation and landmark label ground truth ($y, \mathcal{H}_{1}, \mathcal{H}_{2}$). Eq. \ref{eq:gan_loss} defines our adversarial loss ($\mathcal{L}_{gan}$). Branched network has to enforce its segmentation and also landmark localization to match ground truth labels in order to minimize $\mathcal{L}_{gan}$.
\begin{equation}
	\label{eq:gan_loss}
	\begin{split}
		\mathcal{L}_{gan} = \mathbb{E}_{\textless y, \mathcal{H}_{1}, \mathcal{H}_{2}\textgreater\sim P}[\log D_{pair}(\textless y, \mathcal{H}_{1}, \mathcal{H}_{2}\textgreater)] + \\ \mathbb{E}_{\textless \hat{y}, \mathcal{\hat{H}}_{1}, \mathcal{\hat{H}}_{2}\textgreater\sim G}[1-\log(D_{pair}(\textless\hat{y}, \mathcal{\hat{H}}_{1}, \mathcal{\hat{H}}_{2}\textgreater))]
	\end{split}
\end{equation}
Since the predictions of the branched network at early epochs are rough and can be easily rejected by $D_{pair}$, directly training the branched network from scratch under the adversarial scheme would adversely affect the performance. Thus, we firstly trained the branched network with $\mathcal{L}_{branch}$ to convergence for 16 epochs and then fine-tune it with $D_{pair}$ under the composite loss of $\mathcal{L}_{branch}$ and $\mathcal{L}_{gan}$ for another 5 epochs.

\begin{table*}[!htb] \caption {Comparison of fetal femur segmentation methods} \label{table:Comparison_seg}
	\centering
	\begin{tabular}{c|c|c|c|c|c|c}
		\toprule[2pt]
		Metric		&Unet-ROI		&Unet-S				&BRN 				&BRND 				&BRNDC				&BRNDC-D	\\
		\hline
		DSC[\%]		&87.30$\pm$7.1	&89.41$\pm$4.6		&90.03$\pm$3.5		&90.11$\pm$3.5		&90.76$\pm$3.1		&\textcolor{blue}{91.03$\pm$3.0}\\
		Jacc[\%]	&78.07$\pm$9.8	&81.14$\pm$7.1		&82.04$\pm$5.5		&82.18$\pm$5.6		&83.23$\pm$5.0		&\textcolor{blue}{83.67$\pm$5.0}\\
		Adb[mm]		&1.59$\pm$1.4	&0.84$\pm$0.7		&0.77$\pm$0.5		&0.78$\pm$0.6		&0.68$\pm$0.5		&\textcolor{blue}{0.66$\pm$0.5}\\
		Hdb[mm]		&9.05$\pm$7.4	&4.98$\pm$3.3		&4.69$\pm$3.0		&5.04$\pm$3.4		&4.27$\pm$2.6		&\textcolor{blue}{4.08$\pm$2.7}\\
		Verr[mL]	&5.52$\pm$2.3	&3.04$\pm$1.7		&2.28$\pm$1.2		&1.89$\pm$1.3		&1.79$\pm$1.3		&\textcolor{blue}{1.60$\pm$1.1}\\
		\toprule[2pt]
	\end{tabular}
\end{table*}

\begin{table*}[!htb] \caption {Comparison of landmark localization and length errors} \label{table:Comparison_land}
	\centering
	\begin{tabular}{c|c|c|c|c|c}
		\toprule[2pt]
		Task		&Unet-L 			& BRN 				& BRND 						& BRNDC 			& BRNDC-D	\\
		\hline
		p1[mm] 		&4.32$\pm$3.02		&4.26$\pm$2.64		&4.19$\pm$	2.22			&3.92$\pm$2.64		&\textcolor{blue}{3.70$\pm$1.93}		\\
		p2[mm] 		&4.57$\pm$2.13		&4.46$\pm$2.88		&\textcolor{blue}{4.05$\pm$2.60}		&4.28$\pm$2.78		&4.26$\pm$2.53		\\
		Lerr[mm]  	&1.05$\pm$1.15		&0.96$\pm$1.06		&0.89$\pm$0.99				&1.04$\pm$1.05		&\textcolor{blue}{0.87$\pm$0.91}		\\
		\toprule[2pt]
	\end{tabular}
\end{table*}

\section{Experimental Results}
Our method is validated on fetal femur ultrasound volumes. The dataset contains 50 annotated volumes with size of 416$\times$416$\times$284 and a voxel size of 0.38$\times$0.38$\times$0.38 $mm^{3}$. Approved by local Institutional Review Board, all volumes were anonymized and acquired by an experienced sonographer using a Sonoscope S50 ultrasound machine with an integrated 3D probe. The probe has a $75^{\circ}$ scan angle to ensure a complete scanning of the whole femur. Varying femur poses are allowed in scanning. The dataset covers gestational age from 23 to 31 weeks. An expert with 5-year experience manually delineated all volumes and annotated two femur tips as ground truth for segmentation and localization. 30 volumes are randomly selected for training and the rest as testing. Training set is further augmented with scaling, rotation and flipping to 840. Limited by GPU memory, the femur ROI is rescaled as 0.65 times. There is no pre-alignment for ROIs. Training and testing were run in a NVIDIA GeForce GTX TITAN X GPU (12GB). \par

Segmentation evaluation criteria include Dice Similarity Coefficient (DSC, \%), Jaccard index(Jacc, \%), Average Distance of Boundaries (Adb, $mm$), Hausdorff Distance of Boundaries (Hdb, $mm$) and absolute volume error (Verr, $mL$). Absolute displacement ($mm$) and absolute femur length error (Lerr, $mm$) are used to evaluate the localization of femur tips p1 and p2. Fetal femur length is defined as the Euclidean distance between p1 and p2. Our basic branched network is denoted as BRN. The distance constraint involved version is BRND, the version further equipped with cross connections is BRNDC. The BRNDC trained in the adversarial scheme is denoted as BRNDC-D. Additionally, we implemented two basic U-nets as baselines for fair comparisons. Unet-S is for segmentation. It has the same encoder, decoder and skip connections as the segmentation branch of BRN. Unet-L is for localization. It has the same encoder, decoder and skip connections as the localization branch of BRN. We implemented all the compared methods in 3D fashion with \textit{TensorFlow}.
\begin{figure}[htb]
	\begin{minipage}[b]{1.0\linewidth}
		\centering
		\subfigure[ ]{\includegraphics[width=.46\linewidth]{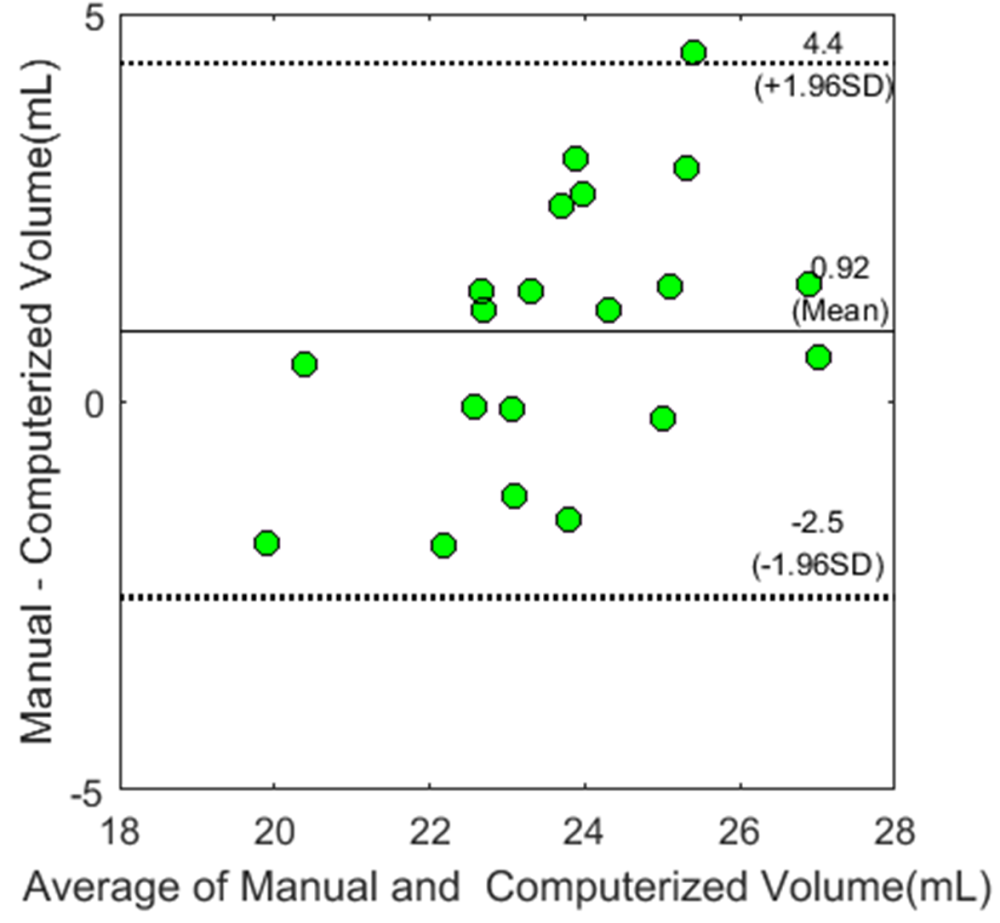}}
		\subfigure[ ]{\includegraphics[width=.46\linewidth]{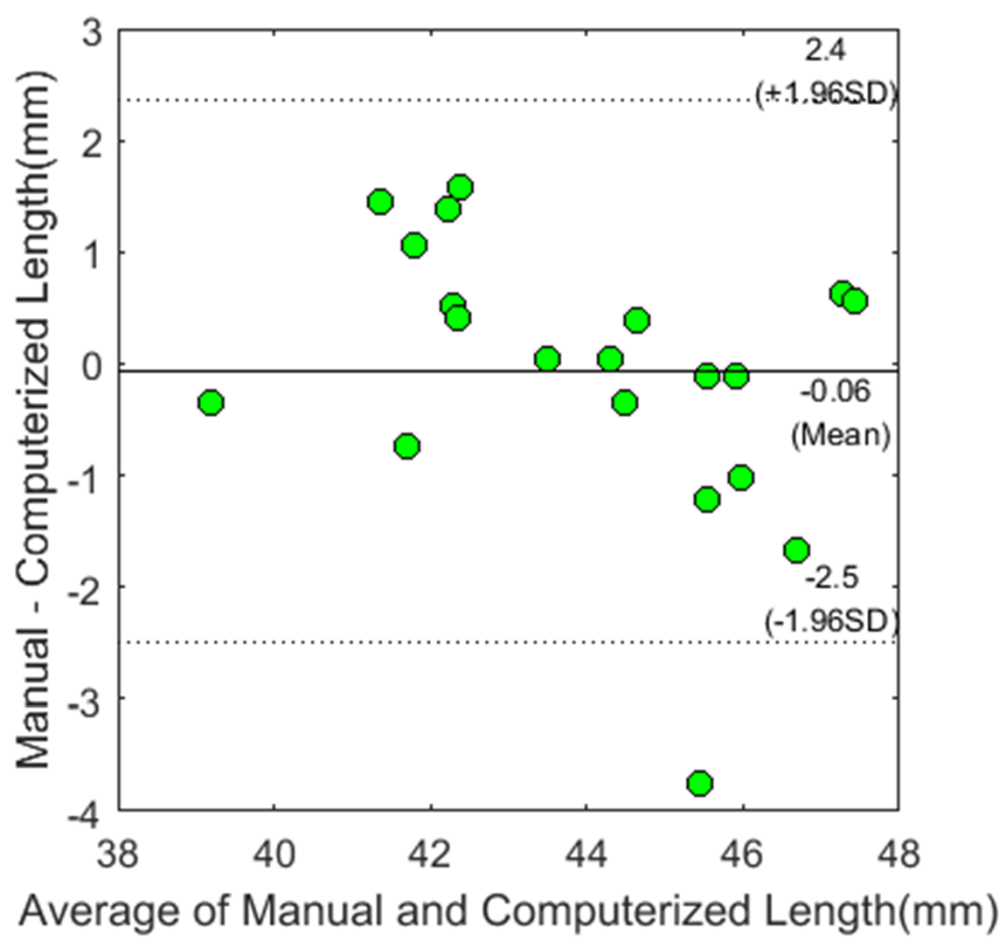}}
	\end{minipage}
	\caption{Bland-Altman plot between BRNDC-D and experts on femur volume measurement (a) and femur length measurement (b).}
	\label{fig:seg_land_bland}
\end{figure}

\begin{figure}[htb]
	\begin{minipage}[b]{1.0\linewidth}
		\centering
		\includegraphics[width=0.90\textwidth]{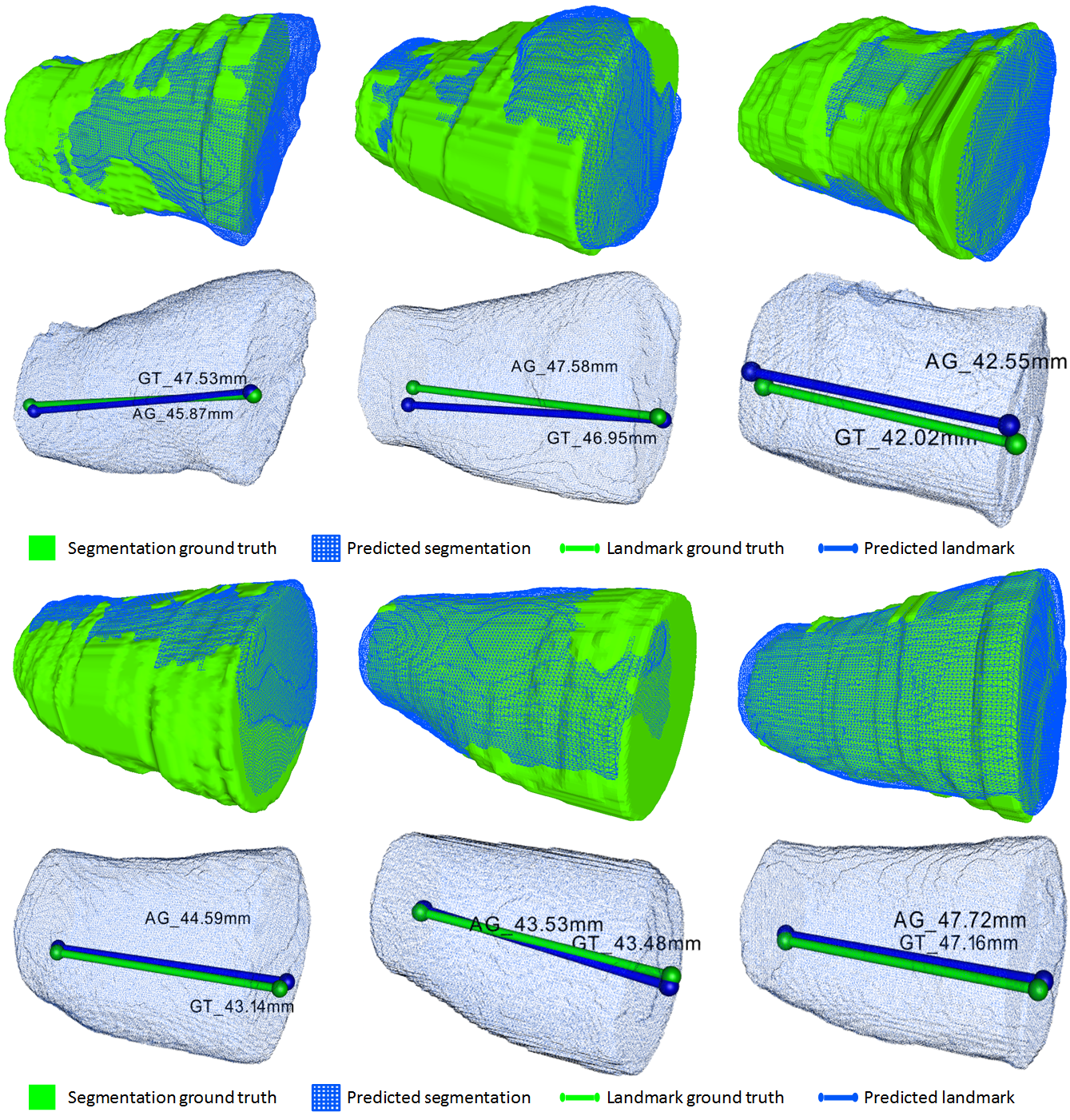}
	\end{minipage}
	\caption{Visualizations of fetal femur segmentation and landmark localization on 6 testing cases. Results from BRNDC-D are compared with ground truth with respect to segmentation, localization and length measurement. Digits stand for femur length, AG for algorithm BRNDC-D, GT for ground truth.}
	\label{fig:seg_visual}
\end{figure}
Quantitative comparisons among methods are shown in Table \ref{table:Comparison_seg} and \ref{table:Comparison_land}. For segmentation, as implemented in 2D whole slice, Unet-ROI performs modest in accurately segmenting the femur among all methods. Our proposed branch architecture and modules successively improve the segmentation over the baseline Unet-S. The largest DSC improvements occur when branch design (BRN, 0.62\%) and cross connections (BRNDC, 0.65\%) are used. Adversarial training further contributes to another 0.3\% improvement in DSC. Distance constraint (BRND) slightly improves DSC, and helps to reduce the volume measurement error for about 13\%. BRNDC-D finally reduces the volume measurement error for about 50 percent. For landmark localization, both branch architecture (BRN) and distance constraint (BRND) promote the localization of two landmarks over Unet-L, especially the p2. Effectiveness of distance constraint is proved and may be extended to other cases, like \cite{huang2018omni}. Although improves the p1 localization and femur segmentation, cross connections (BRNDC) cause slight performance drop in localizing landmark p2. Localization of p2 may be interfered by the segmentation branch via cross connections. Enforcing the correspondence between two branch outputs, adversarial module (BRNDC-D) alleviates the problem and achieves a better balance between segmentation and localization branch. BRNDC-D achieves the best results in segmentation and localization on almost all metrics.

Fig.~\ref{fig:seg_land_bland} shows the Bland-Altman plot to evaluate the agreement between BRNDC-D and experts on fetal femur volume and femur length measurement. As observed, high agreements are achieved in both plots since 95\% of the measurements locate in the $\pm$1.96 standard deviation range. Visualizations of segmentation and landmark localization (Fig. \ref{fig:seg_visual}) produced by BRNDC-D also show good alignment with ground truth. \par

\section{Conclusions}
In this paper, we present an effective framework for simultaneous segmentation and landmark localization in fetal femur ultrasound volumes. Promising segmentation and localization accuracy are achieved on the challenging tasks. We get a good starting point with a branched network to handle these two tasks. Informative cues of segmentation and landmark localization can be propagated bidirectionally through cross connections to benefit each other. The proposed distance based loss and adversarial training scheme suppress the false positives and enhance the localization and segmentation. Our framework is general and has potentials to be extended to similar tasks in volumetric ultrasound. \par

\section{Acknowledgment}
We would like to acknowledge all the volunteers who participated in this research and the expert who made great effort. This work was supported in part by the National Natural Science Foundation of China under Grant 61571304, Grant 81571758, and Grant 61501305, and in part by the National Key Research and Development Program of China under Grant 2016YFC0104703, and a grant from Hong Kong Research Grants Council, under General Research Fund (Project No. 14225616).. \par

\bibliographystyle{aaai}
\bibliography{refs}

\end{document}